\documentstyle[12pt]{article}
\newcommand{\lwig}{\mbox{\raisebox{.3ex}{$<$}$\!\!\!\!\!
$\raisebox{-.9ex}{$\sim$}}}
\newcommand{\gwig}{\mbox{\raisebox{.3ex}{$>$}$\!\!\!\!\!
$\raisebox{-.9ex}{$\sim$}}}
\newcommand{\sigtot}{ \mbox{$\sigma_{\rm total}^{\rm \bar{p} p}$} }
\newcommand{\sigtotweak}
{\mbox{$\sigma_{\rm total}^{\nu_e \bar{\nu}_\mu}$}}
\begin{document}
\baselineskip=18pt

\begin{titlepage}
\title{The Perturbative Onset of Multiparticle \\
Production in Weak Interactions\thanks{This work
was supported in part by a  World Laboratory Fellowship and the
Eloisatron Project.}}
\author{D.A. Morris\thanks{Electronic address : morris@uclahep.bitnet}
and R. Rosenfeld\thanks{Current Address:
Department of Physics, Northeastern University, Boston, MA 02115,
(rosenfeld@neuhep.hex.northeastern.edu)}
\\ {\it Department of Physics}
\\ {\it  University of California Los Angeles}
\\ {\it Los Angeles, California 90024-1547} }
\date{UCLA Preprint 92/TEP/38 \\ October 1992}
\maketitle
\begin{abstract}
We use perturbation theory to estimate
the energy scale beyond which
multiparticle final states become
a dominant feature of high energy weak interactions.
Using estimates from a weak parton model and comparing
two, three and four body final states we deduce that
multiparticle states become important at energy scales
in the range $10^7-10^9$~{\rm GeV}.
\end{abstract}
\end{titlepage}

\section{Introduction}

\hspace*{\parindent}
There have recently been numerous conjectures regarding
spectacular multi-TeV electroweak phenomena,
characterized by the production of many
electroweak gauge and Higgs bosons,
which may be accessible to the next generation of colliders.
Though such phenomena were initially
studied in the context of topologically
induced baryon plus lepton number
violating interactions in the Standard Model
\cite{rigwald-espinosa}, it was
subsequently realized
that they are a general feature of large order tree-level processes,
where dramatic enhancements in production cross sections
may be attributed to phase space and
combinatorial factors\cite{multi-particle}.

Yet another proposal for multiple weak gauge boson production  was
put forward recently by
Ringwald {\it et al.}\cite{ring1,ring2} under the name of
geometrical flavour interactions.
They propose an analogy between multiparticle production
in Quantum Chromodynamics (QCD), which occurs above
energy scales of a few ${\rm GeV}$, and possible
multiparticle production in weak interactions above the
TeV scale, where weak gauge boson masses may be neglected.
More precisely, they relate QCD to a prototypical
confined phase of weak SU(2) and, using scaling arguments,
extrapolate the analogy to the Higgs phase. As a consequence
they allege that above parton-parton center
of mass energies of 2 to 20~TeV,
the non-perturbative production of $O(\alpha_w^{-1}) \approx 30$
electroweak bosons could occur with cross sections of
$O(1$~nb -- $10~\mu{\rm b}).$
Irrespective of the appropriateness of their scaling arguments,
Ringwald {\it et al.} qualitatively suggest that,
as far as weak interactions are concerned,
colliding quarks and leptons may be thought of as being surrounded
by clouds of virtual weak bosons which eventually result
in ``black disc'' cross sections, and hence the name of
geometrical flavour interactions.

        We remark that Ringwald {\it et al.}'s qualitative view
of high energy weak interactions is reminiscent of the QCD parton
model in which an energetic proton may be viewed as a collection of
quarks and gluons which form a cloud of transverse size
$O(\Lambda_{\rm QCD}^{-1}).$
In contrast to the approach of Refs.~\cite{ring1,ring2},
we investigate in this paper the possibility of
high energy weak multiparticle production in the context
of perturbation theory using techniques
similar to those used for multiparticle production in QCD.
We take two successively refined viewpoints.
First, we adopt a weak parton model in which weak gauge bosons
assume the role played by gluons in QCD and look for
the failure of the lowest order approximations to the corresponding
weak parton distribution functions which we interpret as
signalling the increased importance of states with many gauge bosons.
Second, we perform exact calculations of selected $2\rightarrow 2,$
$2\rightarrow 3,$ and $2\rightarrow 4$ weak processes
to investigate particle production at mass scales inaccessible
to the parton model.
For illustrative purposes we concentrate on contributions to the
$\nu_e \overline{\nu}_\mu$ total cross section,
$\sigtotweak$, chosen because it is
naively free from QCD and QED complications.

Our goal is to quantitatively explore the energy range in
which multiparticle final states become a dominant feature
of weak interactions. Since we limit ourselves to the first
few orders of perturbation theory, we will
only consider final states with up to three or four particles
(as opposed to the production
of $O(\alpha_w^{-1})$ particles discussed in Refs.~\cite{ring1,ring2}).
Furthermore, we are interested in the perturbative
decomposition of $\sigtotweak$ into its various multiparticle
channels. The more global question of the
perturbative growth of total (inclusive)
weak cross sections with $\sqrt{s}$ has
been considered recently by Pindor, R\c{a}czka
and Wetterich in Ref.~\cite{pindor} in
which they solve an integral equation suggested by Fadin, Kuraev and
Lipatov\cite{fkl,cheng}. They find slowly
growing weak total cross sections ({\it i.e.}, $O(100~{\rm pb})$
for $\sqrt{s}~ \lwig ~10^{15}~{\rm GeV}$)
but, due to the nature of their methods, they
cannot decompose their results into contributions
with definite numbers of particles in the final state.

Using a variety of techniques we will argue that
weak multiparticle production becomes a dominant characteristic
of weak interactions at energy scales
in the range $10^7-10^9~{\rm GeV}$
which is firmly beyond the reach of any planned accelerator.
Kane and Scanio \cite{kaneandscanio}
have pointed out that cross sections for exclusive processes such
as $e^+ e^- \rightarrow \mu^+ \mu^-$
can be surpassed by higher
order processes (such as $e^+ e^- \rightarrow
\bar{\nu}_e \nu_e \mu^+ \mu^- $ through $W^+W^-$ fusion)
at center of mass energies of O($10^3~{\rm GeV}$).
However this dominance is only relative to
a rapidly falling $s{\rm -channel}$ Born contribution.
In contrast, lowest order $\nu_e \bar{\nu}_\mu$ scattering
proceeds through $t{\rm -channel}$ exchanges which
are analogous to dominant $t{\rm -channel}$ processes in QCD;
at higher orders multiparticle production arises from very
forward ({\it i.e.}, small angle) inelastic reactions.
We shall not consider the
possibility of a strongly interacting Higgs sector which
could, by itself, dominate the characteristics of high
energy weak interactions\cite{chanowitz}.

The paper is organized as follows. In section~2 we
review the QCD parton model approach for describing perturbative
contributions to the antiproton-proton total cross section $\sigtot .$
In section 3 we consider the lowest order distribution
functions of the weak parton model. We use them to
determine contributions to $\sigtotweak$ through
gauge boson fusion and investigate the energy scale at
they become unreliable.
In section~4 we present results of
calculations beyond the weak parton model and argue that
the dominance of multiparticle states at energy scales of
$10^7-10^9~{\rm GeV}$ is more than an artifact of large logarithms
in finite order calculations.

\section{QCD Parton Model and {\bf $\sigtot$}}

\hspace*{\parindent}

Let us abstract those
aspects of the perturbative QCD parton model
of relevance to high energy weak interactions.
Underlying this ambition is the
operating premise, which is not without controversy,
that multiparticle production and
the growth of $\sigtot$ with center of mass energy may
be deduced from the QCD parton model. We briefly
review the support and criticisms of the QCD parton model
view of $\sigtot$ to set the stage
for the weak parton model in the next section.

Though an old idea\cite{clinehalzenluthe},
perturbative contributions to rising hadronic cross sections
gained prominence with the UA1 observation of minijets\cite{ua1papers}
and have subsequently led to many theoretical studies of
the perturbative component of $\sigtot\cite{margolis,durandandpi}.$
However, realistically, the detailed successes of a
perturbative parton model at
momentum transfers at
$O(\Lambda_{\rm QCD})$ must be viewed with a degree
of skepticism since there is little {\it a priori} justification for
a perturbative treatment in that regime. Indeed, descriptions of
$\sigtot$ in terms of the QCD pomeron have also been shown
to accommodate existing data \cite{regge}.

As an approximation to $\sigtot$
in the perturbative QCD parton model
one may write \cite{margolis}
\begin{equation}
\label{eq:qcdfactorization}
\sigtot (s) \simeq \sum_{i,j}
\int_{p_T^2 > 1~{\rm GeV}^2 } dp_T^2
\int dx_1 \, dx_2 \, f_{i/p} (x_1,Q^2) \, f_{j/\bar{p}}(x_2,Q^2)
\, \frac{d \hat{\sigma}_{ij} }{d p_T^2} (Q^2)
\end{equation}
where $Q^2 = x_1 x_2 s,$
$f_{i/p}(x)$ is the number density of parton species $i$
carrying a proton momentum fraction $x$ ({\it i.e.}, a parton
distribution function)
and $\hat{\sigma}_{ij}(Q^2)$ is the
subprocess cross section for colliding partons of species $i$ and $j.$
(It is actually more appropriate to evaluate the distribution functions
at a scale $O( p_T^2 )$  but such details will not concern us.)
The sum extends over all possible quark and gluon species.
The {\it ad hoc} infrared cutoff $p_T^2~\gwig~1~{\rm GeV}^2$
avoids the non-perturbative growth of
$\alpha_s(p_T^2~\rightarrow~\Lambda_{\rm QCD}^2).$

In the parton framework
the perturbative behaviour of $\sigtot (s)$
is accommodated as follows.
As the center of mass energy $\sqrt{s}$ increases,
successively smaller $x$ values become kinematically
accessible. For example,
the small-$x$ behavior of the gluon distribution
function $f_{g/p}(x,Q^2)$ is thought to behave as
$\simeq x^{\delta}$ (where $\delta<0$) so that gluon-gluon
collisions give large
contributions to the total cross section\cite{robertsbook}.
The cooperation of these kinematic and dynamic effects
may be illuminated by a
toy calculation in which one considers only gluon-gluon collisions
in Eq.~\ref{eq:qcdfactorization},
and takes $f_{g/p}(x) \simeq 1/x$ and
$\hat{\sigma}_{gg} \simeq ~$constant,
which gives $\sigma_{\rm tot} \simeq \ln^2 s.$
The enhanced small-$x$ behavior of parton densities inside hadrons
suggests the picture of a high
energy proton as a cloud of small-$x$ partons.
Indeed, using more
 realistic distribution functions and hard cross sections
one can reproduce the Tevatron result for
$\sigtot \simeq~70-80$~mb at
$\sqrt{s} = 1.8~{\rm TeV}$\cite{recdfsigma}.
The ability to mimic this result using perturbative techniques
is based upon the assumption that the total cross section
is dominated by semi-hard
collisions; the infrared cut-off is then interpreted as the scale at
which semi-hard processes become relevant.
In the following section we exploit this apparent
success of perturbative QCD and apply similar
techniques to describe high energy weak interactions where weak
gauge bosons assume the role played by gluons in QCD.

\section{The Weak Parton Model}

\subsection{Factorization and Distribution Functions}
\hspace*{\parindent}
In QED there are already examples of rising cross sections which
may be understood in the context of a parton picture. Among them
is the two-photon production of a fermion
pair $(f \bar f$) through
$e^{+} e^{-} \rightarrow e^{+} e^{-} f \bar{f}$
 which has a lowest order cross section\cite{landau}
\begin{equation}
\sigma( e^{+} e^{-} \rightarrow e^{+} e^{-} f \bar{f} )
=
\displaystyle{ 28 \alpha_e^4 \over 27 \pi m_f^2}
\ln^2 \left( \displaystyle{s \over m_e^2} \right)
\ln \left(   \displaystyle{s \over m_f^2} \right) ,
\label{eq:twophoton}
\end{equation}
which may be obtained by
convoluting the probability of finding quasi-real photons inside
electrons
with the subprocess cross section for fermion
pair production from two on-shell photons.
This is the Weiszacker-Williams
effective photon approximation \cite{epa}, which is
in good agreement with exact calculations\cite{berends}.
The parton model approach to QED has also been successfully applied to
describe other higher order electromagnetic
processes such as corrections to the Z boson line shape\cite{QED}.

The enhanced small-$x$
behaviour of the gluon distribution function in QCD
and the large QED logarithms in Eq.~\ref{eq:twophoton}
have counterparts which we wish to
exploit by considering a parton model for the weak interactions.
By analogy with QCD, we consider writing contributions
to the $\sigtotweak$ in the factorized form
\begin{equation}
\sigtotweak (s) =
\sum_{ i,j}
\int \, dx_1 \, dx_2 \,
f_{i/ \nu_e       }(x_1,Q^2) \,
f_{j/ \bar{\nu}_\mu}(x_2,Q^2) \,
\hat{\sigma}_{ij} (Q^2)
\label{eq:ewkfactorization}
\end{equation}
where $Q^2 = x_1 x_2 s.$
In principle, the sum in Eq.~\ref{eq:ewkfactorization} extends over all
weak parton species found ``inside'' neutrinos (e.g.,
$W,$ $Z,$ $\nu, \cdots $). The weak parton distribution functions
and the on-shell parton-parton subprocess cross sections have
interpretations similar to those in QCD.
At the level of this discussion we emphasize that the
factorization of $\sigtotweak$ in Eq.~\ref{eq:ewkfactorization}
is performed strictly by analogy with QCD and
as such is an
assumption without rigorous justification\cite{collinssoper}.
As in QCD, the scale at which the distribution functions are
evaluated is not actually $x_1 x_2 s$ but rather is determined by the
subprocess under consideration.

In contrast to the QCD factorization,
the integral in Eq.~\ref{eq:ewkfactorization}
does not have an arbitrary infrared cut-off imposed
by the dynamics of the theory.
Instead, there are implicit kinematic constraints of the form
\begin{equation}
\label{eq:weakconstraint}
x_1 x_2 \geq \displaystyle{ ( m_i + m_j )^2 \over  s},
\end{equation}
as required by the
assumption that the weak partons participating
in the hard interaction are on-shell with masses $m_i$ and $m_j.$
In principle, similar constraints are present in QCD but are
usually irrelevant since they are superseded by
the more restrictive constraint on $p_T^2$ associated
the onset of non-perturbative phenomena.

The constraints of Eq.~\ref{eq:weakconstraint} concern
us for the following reasons. Drawing from our
experience with QCD and QED, we anticipate that the
small-$x$ enhancement of the distribution functions
of weak gauge bosons is a promising place to look
for cross section enhancements. However due
to Eq.~\ref{eq:weakconstraint} the
weak parton model can only address gauge boson subprocesses
above energy scales of $O(M_W).$ It is uncertain
beforehand whether the bulk of the multiparticle production
will occur above or below this scale. While multiple gauge
boson production certainly requires energies of $O(M_W)$, the
production of single gauge bosons or light quarks and leptons
can occur
below this scale. For weak interactions we can, unlike QCD,
always perform an exact perturbative calculation (and we do) to check
the sensibility of our assumptions. However we initially
follow the more intuitive route of the weak parton model
to elucidate the operative mechanisms.

The weak parton
distribution functions in Eq.~\ref{eq:ewkfactorization}
obey coupled evolution equations analogous to the
Dokshitzer-Gribov-Lipatov-Altarelli-Parisi\cite{dglap}
equations in QCD and the Lipatov equations in QED\cite{lipatov}.
Due to the relatively small coupling constants involved,
the evolution equations for QED and weak interactions
can be solved iteratively in terms of parton {\it splitting }
functions (see Appendix~A) starting
from well defined input distributions (e.g.,
$f_{\nu / \nu}(x,Q^2 = 0 ) = \delta (1-x)$ ).
In contrast, the strong QCD coupling constant
forbids such an expansion in most regions of interest
and one must solve the evolution equations numerically
starting from input distributions measured, for example,
in deep inelastic scattering experiments\cite{robertsbook}.

Suppose we wish to find the distribution function
of $W^+$ bosons with helicity $\lambda$,
carrying a momentum fraction $x,$
inside an electron neutrino, $ f_{W^+_\lambda / \nu_e}(x,Q^2).$
The lowest order expansion for this distribution function is
\begin{equation}
\label{eq:wplusfunction}
f_{ W^+_\lambda / \nu_e  }(x,Q^2)
 =  P_{ W^+_\lambda / \nu_e  }(x,Q^2)
\end{equation}
where $P_{ W^+_\lambda / \nu_e  }(x,Q^2)$ is the splitting function
 which embodies the details of an elementary
$W^+ \nu_e  e^-$ vertex (see Appendix~A).
Keeping only this first term
of the expansion is known as
the effective vector boson approximation (EVBA)\cite{evba}
which is the weak interaction analog of
the effective photon approximation.
Weak splitting functions
have been derived in many different contexts under
various assumptions\cite{lindfors,rolnick,wuki,kunzt-soper}.
Different authors agree for terms proportional to
$\ln (Q^2/M_W^2)$, the so-called ``leading-log'' contributions,
but usually not further. Additional terms in the splitting functions
can depend on the context of the derivation which can lead to
process dependence and gauge dependence.
If the EVBA is used in the context
of, for example, Higgs production where $Q^2 \gg M_W^2,$
the ``leading-log'' term in the splitting function is, in fact,
the most important. However, our calculations require the use
of more refined splitting functions, listed in Appendix~A,
which are slight variations of those of
Johnson~{\it et al.}\cite{wuki}.

\subsection{Naive Onset of Multiple Gauge Boson Emission}
\hspace*{\parindent}

The effective number of small-$x$ partons associated with
a high energy proton is closely related to
the strength of the QCD coupling constant and the inability
to solve the QCD evolution equations in a iterative manner
starting from first principles.
Likewise, the number of weak partons associated with a
high energy fermion can suggest the suitability
of keeping only the lowest order approximations to the
weak parton distribution functions. Turning this
argument around, the failure of the lowest order approximation
to the weak parton distribution functions suggests
an energy scale above which multiparticle processes become
important.
Neglecting, for the moment, the separate issue of
restrictions which govern the feasibility of
factorizing weak cross sections,
it is amusing to consider two qualitative arguments
leading to an estimate of such an energy scale.

Consider a neutrino entering a reaction
characterized by a center of mass energy $\sqrt{s}.$
If, for example, the hard subprocess in the reaction involves the
emission of a $Z$ boson from the neutrino, the integral of
the $Z$ boson distribution function over all momentum fractions
roughly corresponds to the number of $Z$ bosons, $N_{Z / \nu}(s),$
``inside'' the neutrino
\begin{equation}
\label{eq:ndef}
N_{Z/\nu}(s) = \int_{M_Z^2/s}^{1} dx P_{Z/\nu} (x,s) .
\end{equation}
The lower limit of integration accounts for the $Z$ boson mass and
it is assumed that $\sqrt{s}$ is large enough to accommodate real
$Z$ boson emission.
The precise kinematic limits are not of critical importance
for our argument. In Fig.~1 we plot $N_{Z/\nu}(s)$
which is summed over all boson polarizations although the
transverse polarizations dominate the sum.
It is relatively straightforward to show (see Appendix~A)
that as $s \rightarrow \infty ,$
\begin{equation}
N_{Z/\nu}(s) \simeq
\displaystyle{\alpha_{w} \over 8 \pi \cos^2 \theta_w }
\ln^{2} (s/M_{Z}^{2}) ,
\end{equation}
which becomes of order unity at
$\sqrt{s} \simeq 10^{8} $ GeV.
We interpret this simple estimate as a scale above which the
assumption
of single $Z$ boson emission is inappropriate;
higher order corrections representing the possibility of multiple
emissions should then be considered.

Had we used the ``leading-log'' approximation to the
$Z$ boson distribution function, we would have found
\begin{equation}
N_{Z/\nu}(s) \simeq
\displaystyle{\alpha_w \over 4 \pi \cos^2 \theta_w }
\ln^{2} (s/M_{Z}^{2})
\end{equation}
which is of order unity at $\sqrt{s} \simeq 10^{6} $ GeV.
This lower scale reflects the well known shortcoming of the
``leading-log'' approximation for describing small-$x$ fractions
within the EVBA\cite{wuki}.
For most applications discussed previously
in the literature
this small-$x$ region is
irrelevant but for us it is not.

A second technique to estimate the onset
of multiple gauge boson emission involves a
consideration of the iterative solution
to the evolution equations governing the parton
distribution functions. In QCD this approach forces one
to iterate to all orders because of the
large coupling constant and the singular nature of splitting functions
near
the endpoints $x=0$ and $x=1.$ In the weak
parton model the gauge boson masses avoid splitting function
singularities and the small coupling
suggests the validity of a series expansion.
At asymptotically high energies the weak parton model should
behave more like QCD in the sense that the gauge boson masses
can be neglected. A crude measure of when this occurs
is to ask when the second order term in the weak parton model expansion
of the distribution function becomes comparable to the
first order term.

The second order contribution to the $Z$ distribution function
in terms of first order splitting functions,
depicted in Fig.~2, may be written as
\begin{equation}
f^{(2)}_{Z / \nu} (x,s)
= \int_{x}^{1- M_{Z}^{2}/s} \frac{dy}{y}
\; P_{\nu / \nu} (y,s)  \;
P_{Z / \nu} (x/y,ys)
\end{equation}
where $P_{\nu / \nu} (y,s) = P_{Z / \nu} (1-y,s)$
corresponds to the emission of a $Z$ boson by a neutrino.
It is convenient
to consider the ratio of the second order contribution to the
first order (splitting function) contribution
\begin{equation}
R(x,s) = \displaystyle{
f^{(2)}_{Z / \nu} (x,s)
\over
P_{Z / \nu} (x,s) } .
\end{equation}
The quantity $R(x,s)$ is insensitive to $x$ for small $x,$ where
the distribution functions are largest. The plot of
$R(x=10 M_{Z}^{2}/s,s)$ in Fig.~3 suggests that
the second order correction becomes comparable to the first order term
at $\sqrt{s} \approx 10^{8} $ GeV.

It is pleasing, though perhaps not surprising, to find agreement
between the above two techniques which suggest the onset
of multiple gauge boson emission.
Nevertheless, these arguments must not be taken
too literally since we have ignored the subtleties of whether
weak factorization sanctions such calculations. Indeed, in our
estimation of $\sigtotweak$ in the next section, we will encounter
a context in which we must be considerate of factorization concerns.

\subsection{Parton Model Contributions to {\bf $\sigtotweak$} }

\hspace*{\parindent}

The elastic and inelastic Born cross sections for
$\nu_e \bar{\nu}_{\mu} $ scattering at
center of mass energy
$\sqrt{s}$ are given by
\begin{eqnarray}
\sigma_{\rm elas}^{\nu_e\bar{\nu}_\mu}  (s)
& = &
\displaystyle{\pi \alpha_w^2 \over 4 \cos^4 \theta_w M_Z^2}
\left[ 1 + 2 z
- 2 z ( 1 + z ) \log \left( \displaystyle{ 1 + z \over z } \right)
\right]
\label{eq:elastic} \\
\sigma_{\rm inel}^{\nu_e\bar{\nu}_\mu} (s)
& = &
\qquad \, \, \displaystyle{\pi \alpha_w^2 \over M_W^2}
\left[ 1 + 2 w
- 2 w ( 1 + w ) \log \left( \displaystyle{ 1 + w \over w } \right)
 \right]
\label{eq:inelastic}
\end{eqnarray}
where $w = M_W^2 / s$ and $z = M_Z^2 / s.$ For
$\alpha_w^{-1} \simeq 30,$ $M_Z = 91.17~{\rm GeV}$ and
$M_W = 80~\rm{GeV}$ these expressions for
$\sigma_{\rm elas}^{\nu_e\bar{\nu}_\mu}  (s)$
and
$\sigma_{\rm inel}^{\nu_e\bar{\nu}_\mu} (s)$
respectively
asymptote to $\simeq 70~{\rm pb}$ and $\simeq 210~{\rm pb}$ for
$\sqrt{s}~\gwig~1~{\rm TeV}.$

In order to estimate the energy at which
$\sigtotweak$ may begin to receive
significant contributions from multiparticle
production,
consider the gauge invariant
set of diagrams of Fig.~4 corresponding
to the exclusive production of an additional fermion pair
in the reaction
$\nu_e \bar{ \nu}_\mu \rightarrow e^- \mu^+ f \bar{f}.$
We shall focus on this process as being representative of a class of
reactions
at this order of perturbation theory, though many other final states
are certainly possible. Let us consider how these diagrams
fit into a distribution function framework.

In analogy with QCD, the electroweak factorization ansatz of
Eq.~\ref{eq:ewkfactorization}
associates the diagrams of Figs.~4a,b with
a $W^+ W^-$ fusion subprocess. Included in Fig.~4c are
final state bremsstrahlung processes
corresponding to fragmentation functions
of the final leptons and initial state bremsstrahlung processes
related to redefinitions of the neutrino structure functions.

We can accommodate the $2 \rightarrow 2$ Born processes
and the processes of Figs.~4 in the parton framework by
summing only over $W^+W^-$
and $\nu_e \bar{\nu}_\mu$ subprocesses in
the factorized expression
for $\sigtotweak$ in Eq.~\ref{eq:ewkfactorization}.
In addition, we only keep terms up to $O(\alpha_w)$ in
$f_{ W^+_\lambda / \nu_e  }(x,Q^2)$
and $f_{ W^-_\lambda / \bar{\nu}_\mu }(x,Q^2)$ as in
Eq.~\ref{eq:wplusfunction}, and in
\begin{eqnarray}
f_{ \nu_e / \nu_e }(x,Q^2) & = & \delta ( 1 - x ) +
P_{Z / \nu_e }(1-x,Q^2),  \\
f_{ \bar{\nu}_\mu / \bar{\nu}_\mu }(x,Q^2)
& = & \delta ( 1 - x ) + P_{Z / \bar{\nu}_\mu }(1-x,Q^2),
\end{eqnarray}
where the delta functions reflect the initial conditions
of the neutrinos.
Consequently, in the factorization approximation, we use the above
expansions to obtain
\begin{equation}
\sigtotweak(s)
=
   \sigma_{\rm elas}^{\rm Born} (s)
 + \sigma_{\rm inel}^{\rm Born}(s)
 + \sigma_{W^+W^-}^{\rm fusion}(s)
 + \sigma^{\rm brem}(s)
 + \cdots
\end{equation}
where the explicitly listed terms correspond to particular
subprocesses which we will discuss in turn.

The cross sections $\sigma_{\rm elas}^{\rm Born} (s)$ and
$\sigma_{\rm inel}^{\rm Born}(s)$ are given by
Eqs.~\ref{eq:elastic},\ref{eq:inelastic}.
The $W^+W^-$ fusion cross section is given by
\begin{eqnarray}
\label{eq:fusion}
\lefteqn{ \sigma_{W^+W^-}^{\rm fusion}(s)   =  }  \\ \nonumber
& & \sum_{\lambda , \lambda '} \int_{0}^{1} dx_1 \,  \int_{0}^{1} dx_2
\,
P_{ W^+_{\lambda  }/ \nu_e} (x_1, s)
P_{ W^-_{\lambda '}/ \bar{\nu}_{\mu}} (x_2, x_1 s)
\hat{\sigma}_{ W^+_{\lambda  }
                    W^-_{\lambda '} \rightarrow
                    f \bar{f} } (x_1 x_2 s)  \\ \nonumber
& & \qquad \qquad \qquad \qquad \qquad \times
\Theta (x_1 x_2 s - 4 M_W^2)
\end{eqnarray}
where the sum extends over the polarization states of
the on-shell $W$ bosons. The subprocess cross sections for
$ W^+_{\lambda} W^-_{\lambda '} \rightarrow
f \bar{f}$ are straightforward to calculate and
may be found in the literature\cite{gaemers}.
The kinematic constraints of the parton model strictly limit the
applicability of this particular contribution
to fermion pair masses $m_{\bar{f} f} \geq 2 M_W.$
The bremsstrahlung contributions $\sigma^{\rm brem}$
correspond to the emission of a near-mass-shell photon or
$Z$ which subsequently decays into a $f \bar{f}$ pair;
since these only contribute to
$m_{\bar{f} f} < 2 M_W$
(which falls outside the constraints of Eq.~\ref{eq:fusion})
we neglect them in this approximation.

In Fig.~5b we show
$\sigma_{W^+W^-}^{\rm fusion}$
summed over all
lepton and quark species (except for $t$ quarks)
such that $m_{f \bar{f}} \geq 2 M_W.$
The $\ln^2 s$ growth of this contribution to
$\sigtotweak$ originates from the logarithms
in the splitting functions corresponding to finding transverse
gauge bosons inside neutrinos, in complete analogy with
the QED expression of Eq.~\ref{eq:twophoton}. In the next
section we compare this result with an exact calculation.

\section{Multiparticle Production Beyond the Parton Model}

\hspace*{\parindent}
Fig.~5c gives the result of an exact calculation
for fermion pair production for
$m_{\bar{f} f} \geq 2 M_W$ which we performed using
helicity techniques\cite{helicityref}.
The helicity calculation
incorporates interference effects between the diagrams
of Fig.~4 and grows as $\ln s.$
As discussed in more detail in
Refs.~\cite{kunzt-soper,kleiss-stirling} the fact
that the parton model behaviour
disagrees with the exact calculation only by a power of $\ln s$
and not by some power of $s$ (which would be characteristic
of a violation of gauge invariance in the EVBA)
is somewhat of a success for the EVBA which neglects the effects of
bremsstrahlung diagrams. Furthermore, closer inspection reveals that
in the weak parton model the dominant contribution to $f \bar{f}$
production originates from the threshold region
($m_{W^+ W^-} = m_{f \bar{f}}~\gwig ~2 M_W$) whereas the
validity of the EVBA to $W^+W^-$ fusion requires that
$m_{W^+W^-} \gg 2 M_W.$
It is at this level of agreement (or disagreement) that the weak parton
model is useful for calculating low order contributions to
$\sigtotweak .$
In any case, the growing contribution to $\sigtotweak$ from
fermion pair masses above $2 M_W$ is not significant when
compared with the much larger Born terms.

Fortunately, we are not limited to using the weak parton
model exclusively. We next consider
particle production processes below the $2 M_W$ threshold
of the weak parton model. By integrating our exact
$2\rightarrow 4$ calculation over all
fermion pair masses we obtain a contribution to $\sigtotweak$ of
$O(\alpha_w / M_W^2 \ln (s/M_W^2)),$
as shown in Fig.~5a, which is dominated by the
$Z$ boson pole. The same behavior is more easily deduced
from the $2 \rightarrow 3$ production of an on-shell
$Z$ boson (see Fig.~6)
which agrees with the $2 \rightarrow 4$ calculation
(except that the latter has an additional $O(\alpha_w^4)$ component
corresponding to the region $ | m_{\bar{f} f} - M_Z | > \Gamma_Z )$.

In Fig.~7 we plot additional contributions
to $\sigtotweak$ from the production of on-shell $W^\pm$
and Higgs bosons (treating them as stable particles).
At $\sqrt{s} \simeq 10^{7} ~{\rm GeV}$
the sum of the $2 \rightarrow 3$ contributions
is approximately half that of $\sigma^{\rm Born}_{\rm elas} +
\sigma^{\rm Born}_{\rm inel} \simeq 290 ~{\rm pb}.$
To be consistent, we should more appropriately compare
the $2 \rightarrow 3$ contributions
with the radiatively corrected $ 2 \rightarrow 2$ final
states.
To do this we consider the
limit of the standard electroweak model
without QED
({\it i.e.}, $\sin \theta_w = 0$) and take
$M_W = M_Z = 91.17~{\rm GeV}.$ In Fig.~8 we show the sum of the
analogous
$ 2 \rightarrow 3$
contributions as well as an estimate of the of the radiatively
corrected
$ 2 \rightarrow 2$ contributions deduced from the work of
Frankfurt and Sherman\cite{frankfurtandsherman}.

For the $2\rightarrow2$
contribution to $\sigtotweak$ in Fig.~8 we take
\begin{equation}
\sigma_{2 \rightarrow 2} \simeq
\sigma_{\rm elas}^{\nu_e \bar{\nu}_\mu}(s) +
\sigma_{\rm inel}^{\nu_e \bar{\nu}_\mu}(s)
- \Theta( s - s_0 )
\displaystyle{ 15 \pi \alpha_w^3 \over 16 M_Z^2 }
\ln \left(\displaystyle{s \over s_0} \right)
\end{equation}
where $\sigma_{\rm elas}^{\nu_e \bar{\nu}_\mu}(s)$
and $ \sigma_{\rm inel}^{\nu_e \bar{\nu}_\mu}(s)$
are given by Eqs.~\ref{eq:elastic},\ref{eq:inelastic}
and the coefficient of the logarithm may be extracted, with
the help of the optical theorem, from the leading log
calculations of Ref.~\cite{frankfurtandsherman}. As an
additional approximation which fixes the nonleading logarithmic
contribution to the radiative correction, we include the
theta function and adjust the argument of the logarithm
so that the radiative corrections only contribute above
a center of mass energy $\sqrt{s_0}.$
In this approximation the $2 \rightarrow 3 $ contributions
surpass the $2 \rightarrow 2 $ contributions to
$\sigtotweak$ in the neighbourhood of $10^7~{\rm GeV}$
for $\sqrt{s_0}$ in the range from $M_Z$ to $10~{\rm TeV}.$

The appearance of large logarithms in
both the lowest order
$2 \rightarrow 3$ processes and
the next to leading order $2 \rightarrow 2$ processes
raises concerns over the contributions of higher orders of perturbation
theory. As a step towards estimating the effects of higher orders we
adopt a conjecture of Fadin, Kuraev and Lipatov of Ref.~\cite{fkl}
whereby we modify our $2 \rightarrow 2$ and
$2 \rightarrow 3$ amplitudes in an attempt to
sum leading logarithms to all orders.
For the $2 \rightarrow 3$ amplitudes, this
corresponds to recognizing the dominance of the
multiperipheral diagrams (e.g., Fig.~6a) and replacing the
the vector meson propagators with those of reggeons.
We should emphasize that unlike Ref.~\cite{fkl},
it is not clear that such replacements are sensible
in our context. With this caveat in mind, we will show how
our results are modified before discussing their significance.

Consider the contribution to the $2 \rightarrow 3$ amplitude
from the diagram in Fig.~6a. Due to vector boson propagators
this diagram is proportional to
$ (q_1^2 - M^2)^{-1} (q_2^2 - M^2)^{-1}$
with momentum transfers $q_1 = p_{\nu_e} - p_{e^{-}}$
and $q_2 = p_{\bar{\nu}_\mu} - p_{\mu^{+}}.$
In the ansatz of Fadin, Kuraev and Lipatov these propagator factors
are replaced by
\begin{equation}
\displaystyle{ 1 \over  q_1^2 - M^2}
\displaystyle{ 1 \over q_2^2 - M^2 }
\rightarrow
\displaystyle{ \left( s_1 / M^2 \right)^{\alpha(q_1^2)}
\over q_1^2 - M^2 }
\displaystyle{ \left( s_2 / M^2 \right)^{\alpha(q_2^2)}
\over q_2^2 - M^2 }  ,
\end{equation}
where
\begin{equation}
s_1 = ( p_{e^{-}} + p_Z )^2,
\quad \quad
s_2 = ( p_{\mu^{+}} + p_Z )^2 ,
\end{equation}
and
\begin{equation}
\alpha(q^2) =
\displaystyle{ g^2  \over (2 \pi)^3 }
\int dk_\perp^2 \displaystyle{ q^2_\perp - M^2 \over
( k_\perp^2 - M^2 )
( ( q - k )^2_\perp - M^2 )  } \quad .
\end{equation}
Likewise, the $2 \rightarrow 2$ amplitudes are modified
by an overall factor of $(s/M^2)^{\alpha(q^2)}.$ Incorporating these
modifications in our helicity calculations leads to the
cross sections shown in Fig.~9. The $2 \rightarrow 3$
curves in Fig.~9 are extrapolated beyond $10^7~{\rm GeV}$
since our unoptimized numerical integration
becomes unreliable beyond that energy. Nevertheless, in this
approximation we see that
the $2 \rightarrow 2$ and $2 \rightarrow 3$ contributions
to $\sigtotweak$ become
comparable in the vicinity of $10^{10}~{\rm GeV}.$
Again, these results are appropriate for $\sin \theta_w = 0$
and $M = M_W = M_Z = 91.17~{\rm GeV}.$

As mentioned above, the conjecture of Fadin, Kuraev and Lipatov
deserves scrutiny in our context. The implied modification of the
$2\rightarrow 2$ amplitude most
plausibly accounts only for the exchange of weak isospin
$T=1$ in the $t-{\rm channel}$ without manifestly accounting
for the more important $T=0$ (pomeron) exchange.
On the other hand, our estimate of the radiative corrections
to $2 \rightarrow 2$ in Fig.~8 contains both $T=0$ and $T=1$
contributions
(but is only of first order in $\alpha_w \ln s$).
Hence it is perhaps safer to
focus on the qualitative nature of the higher order corrections
suggested in Fig.~9. Namely, that while higher order corrections
tend to suppress the contributions of individual channels, the
energy scale at which $ 2 \rightarrow 3$ reactions
surpass $ 2 \rightarrow 2$ reactions is not extensively affected
(to logarithmic accuracy).

Though we have only explicitly dealt with $ 2 \rightarrow 2,$
$ 2 \rightarrow 3,$ and, to some extent, $2 \rightarrow 4$
contributions to $\sigtotweak$, it is tempting to speculate
beyond few body final states. Fig.~9 suggests that
radiative corrections to $2 \rightarrow 3$ processes
subdue the growth of those channels and may eventually
lead to overall decreasing contributions. Meanwhile
Pindor {\it et al.}\cite{pindor} suggest a slowly
growing $\sigtotweak$  which implies that $2 \rightarrow n$
($n > 3 $) channels must be opening up
to compensate for the decreasing contributions
from $2 \rightarrow 2$ and $2 \rightarrow 3$ channels.
This picture of multiparticle production
is analogous to what is observed experimentally
in QCD, albeit at a much different energy scale.

In summary, we have shown that by considering the lowest
order weak parton model and $2 \rightarrow 2$
and $2 \rightarrow 3$ contributions to $\sigtotweak$ that
multiparticle production may become an important
feature of weak interactions at energy scales of $10^7-10^9~{\rm GeV}.$
In each case the appearance of large logarithms plays an important
role yet we suspect that the relevant energy scale is not an
artifact of a finite order of perturbation theory.

\section*{Acknowledgements}
\hspace*{\parindent}
We would like to acknowledge helpful conversations with
U.~Baur, M.~Cornwall,S.~Khlebnikov, F.~Olness, R.D.~Peccei, A.~Stange
and W.-K.~Tung during the course of this work. D.A.~Morris is supported
by the Eloisatron Project and R.~Rosenfeld is supported
by a World Laboratory Fellowship.

\newpage

\section*{Appendix A}
\hspace*{\parindent}
For completeness, we include in this appendix
explicit expressions for the
vector boson splitting functions used in the text.
The following formulae are essentially those of Johnson {\it et al.}
from Ref.~\cite{wuki} with some minor but notable modifications.

Consider a fermion entering a reaction with a body $A$
at center of mass energy $\sqrt{s}$ (see Fig.~A1).
The splitting function formalism addresses processes
which correspond to the emission of a vector boson
by the fermion so the vector boson is considered
as an initial state particle in a subprocess
with the body $A.$
If the lagrangian coupling between vector bosons and fermions is
$\overline{\Psi} \Gamma_\mu \Psi V^\mu$, where
\begin{equation}
\Gamma_\mu = g_R \, \gamma_\mu \, \displaystyle{ 1 + \gamma_5  \over 2 }
           + g_L \, \gamma_\mu \, \displaystyle{ 1 - \gamma_5  \over 2 } ,
\end{equation}
then the longitudinal vector boson  splitting function is
\begin{eqnarray}
\label{eq:long}
P_{ V_L / f } (x,s) & = &
F_s \left( \displaystyle{ g_L^2 + g_R^2 \over 16 \pi^2 } \right)
\left[ \displaystyle{ 2 ( 1- x)(r+x)\over x ( x-r )} \right. \\
& &  \qquad \qquad
\left. - \displaystyle{ 2 r ( 2 + r - x ) \over ( x- r)^2 }
 \ln \left( \displaystyle{ ( 1 + r - x ) x \over r } \right)
\right] \nonumber
\end{eqnarray}
where $ r = M_V^2 / s $ accounts for the vector boson mass $M_V.$
The spin factor $F_s~=~2$ for neutrinos (and $F_s~=~1$ otherwise)
compensates for the fact that the formulae of Ref.~\cite{wuki}
are spin averaged assuming the initial fermion has two helicity
states.

The splitting functions correspond to the emission of a positive or
negative helicity boson are, respectively,
\begin{eqnarray}
P_{ V_+ / f } (x,s) & = & F_s \left[ g_L^2 h_1(x,r) + g_R^2 h_2(x,r)
\right] \\
  & & \nonumber \\
P_{ V_- / f } (x,s) & = & F_s \left[ g_R^2 h_1(x,r) + g_L^2 h_2(x,r)
\right]
\end{eqnarray}
where
\begin{eqnarray}
\label{eq:h1}
h_1(x,r) & = &
\displaystyle{ 1 \over 16 \pi^2 }
\left[
-  \displaystyle{ ( 1 - x ) ( 2 + r - x ) \over ( x - r ) }
- ( x - r) \ln (x)  \right. \\
\nonumber
&  & \qquad \qquad \left.
+~ (1 + r - x ) \left( \displaystyle{ x + r \over (x - r)^2 } - 1
\right)
\ln \left( \displaystyle{ ( 1 + r - x ) x \over r } \right) \right] \\
  & &  \nonumber \\
h_2(x,r) & = &
\displaystyle{ 1 \over 16 \pi^2 } \left[
- ~\displaystyle{ ( 1 - x ) ( 2 + r - x) \over
( 1 + r - x ) ( x - r ) }
+ \displaystyle{ x + r \over ( x - r )^2 }
\ln \left( \displaystyle{ (1+r-x) x \over r } \right) \right]
\nonumber \\
\label{eq:h2}
  &  &
\end{eqnarray}

Eqs.~\ref{eq:long},\ref{eq:h1},\ref{eq:h2} are slightly different
from the corresponding expressions of Ref.~\cite{wuki}.
Our use of an on-shell flux factor has the effect of
replacing an overall factor of $x$ in the equations of Ref.~\cite{wuki}
with a overall factor of $x-r.$ A consequence of this modification
is that the distribution functions vanish at $x=r.$

The quantity $N_{V/f}(s)$ very roughly corresponds to the
probability of finding a vector boson (of any polarization)
``inside'' a fermion. It is defined as the integral
over $x$ of the vector boson distribution functions, which
to lowest order, is given by

\begin{equation}
N_{V/f} (s) = \int^1_r \, dx \, \left[   P_{V_L / f }(x,s) +
                                  P_{V_+ / f }(x,s) +
                                  P_{V_- / f }(x,s)
                         \right] .
\end{equation}

Using the above expressions for the splitting functions results
in
\begin{equation}
N_{V/f} (s) =
F_s \left( \displaystyle{ g_L^2 + g_R^2 \over 16 \pi^2 } \right)
\left[ \ln^2 r + \displaystyle{7\over 2} \ln r
         +\displaystyle{r^2\over 4} - 5 r + \displaystyle{19 \over 4 }
\right] .
\end{equation}

\newpage
\section*{Figure Captions}

\begin{description}

\item[Figure 1.] Integral of the lowest order $Z$ boson distribution
function for a neutrino.

\item[Figure 2.] Kinematics relevant to second order contribution to
the Z boson distribution function for a neutrino. $s$ is
the invariant mass squared of the original $\nu A$ system.
One $Z$ boson carries away a momentum fraction $1-y$
whereas the $Z$ boson entering the hard subprocess carries a
fraction  $x$ of the original neutrino momentum.

\item[Figure 3.] Ratio of second order contribution to lowest order
Z boson distribution function for a neutrino.

\item[Figure 4.] Gauge invariant set of diagrams for the production
of an additional fermion pair (where $f \neq \mu,e$).

\item[Figure 5.] (a):
$\sigma( \nu_e \bar{ \nu}_\mu \rightarrow e^- \mu^+ Z )$
calculated exactly at tree level;
(b) $\sigma(\nu_e \bar{ \nu}_\mu \rightarrow e^- \mu^+ f \bar{f})$
in effective vector boson approximation $( m_{f \bar{f} }
\ge 2 M_{W} $);
(c) $\sigma(\nu_e \bar{ \nu}_\mu \rightarrow e^- \mu^+ f \bar{f})$
calculated exactly at tree level ($m_{f \bar{f} } \ge 2 M_{W} $);

\item[Figure~6.]
Diagrams for the production of an on-shell $Z$ boson.
(a) Dominant multiperipheral contribution (b) Subdominant
bremsstrahlung diagrams.

\item[Figure~7.]
$2 \rightarrow 3$ contributions to
$\sigtotweak$ in the standard model
with $M_Z=91.17~{\rm GeV}, M_W=80~{\rm GeV}$ and $M_H=500~{\rm GeV}.$
Cross sections for final states containing a Higgs boson
($ \sigma( \nu_e \bar{\nu}_\mu H ) +
\sigma( e^- \mu^+ H )$), W boson
($\sigma( \nu_e \mu^+ W^- )$ + $\sigma( e^- \bar{\nu}_\mu W^+ )$),
Z boson ($\sigma( e^- \mu^+ Z )$ + $\sigma( \nu_e \bar{\nu}_\mu Z)$)
and their sum.

\item[Figure~8.]
Contributions to $\sigtotweak$
in the limit $\sin\theta_w = 0,$ $M_W=M_Z=M_H=91.17~{\rm GeV}.$
Sum of $2 \rightarrow 2$ contributions contains estimated lowest
order radiative corrections which are assumed to vanish below
below $\sqrt{s}=1~\rm{TeV}.$ $2 \rightarrow 3$ curve
is sum of  exact tree level calculation for states containing final
state Z, W or H bosons.

\item[Figure~9.]
Contributions to
$\sigtotweak$
in the limit $\sin\theta_w = 0, M_W=M_Z=M_H=91.17~{\rm GeV}$
with amplitudes modified by the Fadin, Kuraev and Lipatov (FKL)
conjecture described in the text. (a) Sum of $2 \rightarrow 2$ Born
contributions, (b) Sum of FKL modified
$ 2 \rightarrow 2$ contributions,
(c) Sum of $ 2 \rightarrow 3 $ Born contributions,
(d) Sum of FKL modified $2 \rightarrow 3 $ contributions.
The dashed portions of (c) and (d) are extrapolations.

\item[Figure~A1.]
Diagram corresponding the process $f A \rightarrow f' X$
in which the initial fermion $f$ emits a gauge boson $V$
which, in the effective vector boson approximation, is
considered to be on-shell in the subprocess $A V \rightarrow X.$

\end{description}
\end{document}